\documentclass[pra,twocolumn,superscriptaddress,sort&compress]{revtex4-2}
\usepackage{amsmath,amssymb,mathrsfs}
\usepackage{verbatim}
\usepackage{graphicx}
\usepackage{dcolumn}
\usepackage{bm}
\usepackage{float}
\usepackage{appendix}
\usepackage{paralist}
\usepackage{subfigure}
\usepackage[colorlinks=true,linkcolor=red,anchorcolor=blue,citecolor=blue]{hyperref}
\newcommand{\n}{ \notag \\ }
\begin{document}
	\title{ Tomonaga-Luttinger liquid  theory for one-dimensional   attractive Fermi gases }
	
	\author{Hai-Ying Cui}
	\affiliation{Innovation Academy for Precision Measurement Science and Technology, Chinese Academy of Sciences, Wuhan 430071, China}
	\affiliation{University of Chinese Academy of Sciences, Beijing 100049, China.}

	\author{Yu-Hao Yeh}
	\affiliation{Department of Physics and Astronomy,  Rice University, Houston, Texas 77251-1892, USA}
	
	\author{Aashish Kafle}
	\affiliation{Department of Physics and Astronomy,  Rice University, Houston, Texas 77251-1892, USA}
	
	\author{Randall G. Hulet}
	\affiliation{Department of Physics and Astronomy,  Rice University, Houston, Texas 77251-1892, USA}
	
	\author{Han Pu}
	\email[]{hpu@rice.edu}
	\affiliation{Department of Physics and Astronomy,  Rice University, Houston, Texas 77251-1892, USA}
	
	\author{Thierry Giamarchi}
	\affiliation{Department of Quantum Matter Physics, University of Geneva, Geneva, Switzerland}

	\author{Xi-Wen Guan}
	\email[]{xiwen.guan@anu.edu.au}
	\affiliation{Innovation Academy for Precision Measurement Science and Technology, Chinese Academy of Sciences, Wuhan 430071, China}
	\affiliation{Hefei National Laboratory, Hefei 230088, People’s Republic of China}
	\affiliation{Department of Fundamental and Theoretical Physics, Research School of Physics, Australian National University, Canberra, ACT 0200, Australia}

	\begin{abstract}
		
		The one-dimensional (1D) Yang-Gaudin model--an integrable $\delta$-function interacting Fermi gas, serves as a paradigm in quantum many-body physics, encompassing phenomena from spin-charge separation to the Luther-Emery liquid.
		However, a consistent description of the Luther-Emery liquid and the Fulde-Ferrell-Larkin-Ovchinnikov (FFLO)-like pairing states in the 1D attractive Fermi gas remains elusive.
		In this work, we develop a universal Tomonaga-Luttinger liquid (TLL)  theory  to describe the FFLO state both for the case of weak and strong binding between fermions of opposite spins. 
		We rigorously derive a low-energy effective Hamiltonian using bosonization, revealing the emergence of a two-component TLL: one exhibiting spin-charge coupling in the weak binding regime, and another featuring charge-charge separation in the strong binding regime.
		For the weak binding regime, we  further derive renormalization-group equations for the sine-Gordon term in the spin sector and show that this term undergoes a relevant-irrelevant phase transition driven by the magnetic field.
		We further analyze the dynamical correlation functions of the FFLO pairing state based on the effective Hamiltonian we constructed.
		Finally, we discuss  experiments using ultracold atoms to elucidate the Luther-Emery liquid behavior and the subtle phenomena of spin-charge coupling and charge-charge separation.
	\end{abstract}
	
	\maketitle
	\section{I. Introduction}
	
	Interacting quantum many-body systems exhibit novel  phenomena  rooted in   the microcosmic  interactions and excitations among  the  constituent particles.
	Understanding these phenomena has posed major challenges ever since the discovery of quantum physics.
	Numerous methods have been developed to study the universal low-energy physics, including Landau’s Fermi liquid theory \cite{Mahan:1981, Pines:2018}, the density matrix renormalization group (RG) \cite{Schollwock:2011,Solyom:1979,Kollath:2005,Fiete:RMP2007}, and the Green’s function approach \cite{Economou:2006}.
	Notably, the Tomonaga-Luttinger liquid (TLL) theory \cite{Tomonaga:1950,Luttinger:1963,Haldane:1981,Meden:1992,Voit:1993,Giamarchi:2004,Giamarchi:1988,Schulz:1998,Cazalilla:2011,Gogolin:1999,Nagaosa:1999,Tognetti:2000,Schoeller:1998,Schulz:1998} offers a universal framework to describe the low-energy physics of one-dimensional (1D) quantum many-body systems.
	
	Compared to quasiparticle behavior manifested in higher-dimensional systems, excitations and correlations in 1D quantum systems differ significantly, leading to unique phenomena such as collective motion of bosons and edge singularity in the spectral function.
	In 1D interacting systems \cite{Giamarchi:2004,Giamarchi:1988,Schulz:1998}, the low-energy excitations are no longer described by Landau quasiparticles dressed by interactions.
	Instead, within the framework of the TLL theory, they split into two collective modes carrying separately spin or charge, known as spin-charge separation.
	In this context, the TLL description of low-energy physics in 1D repulsive Fermi gases is extensively studied via the bosonization technique \cite{Tomonaga:1950,Luttinger:1963,Haldane:1981,Meden:1992,Voit:1993,Giamarchi:2004,Giamarchi:1988,Schulz:1998,Cazalilla:2011,Gogolin:1999,Nagaosa:1999,Tognetti:2000,Schoeller:1998,Schulz:1998,Gogolin:1999,Nagaosa:1999,Tognetti:2000,Schoeller:1998}.
	Numerous properties have been obtained, including correlation functions, RG analysis \cite{Giamarchi:2004,Giamarchi:1988}, nonlinear TLL theory \cite{Glazman:2009,Egger:2001,Fiete:RMP2007}, and spin-charge coupling induced by external fields \cite{Hideo Aoki:1996} or high-energy interactions \cite{Eran Sela:2010}.
	Spin-charge separation has been experimentally investigated in various quasi-1D platforms \cite{Bouchoule:2025}, including solid-state materials with momentum-resolved tunneling \cite{Auslaender:2002,Auslaeder:2005,Jompol:2009} or angle-resolved photoemission spectroscopy
	\cite{Segovia:1999,Kim:1996,Kim:2006}.
	Spin and charge fractionalization was recently observed in experiments with ultracold atoms in single-site-resolved 1D Hubbard chains \cite{Hilker:2017,Vijayan:2020}.
    Dynamic structure factors(DSFs) of the 1D repulsive Fermi gas have been measured using Bragg spectroscopy\cite{Yang:PRL2018} for both the charge and spin excitations \cite{Hulet:2022,Cavazos:NatComm:2023}, thus enabling a precise test of spin-charge separation.
	
	
	On the other hand, 1D attractive fermions exhibit novel pairing states, whose physical properties strongly depend on polarization. The fully paired case--that is, in the absence of an external magnetic field with zero polarization, forms a Luther-Emery liquid  \cite{Luther:1974}.
	In this state, low-energy excitations are gapped in the spin sector but remain gapless in the charge sector.
	Although the Luther-Emery liquid has been extensively studied \cite{Seidel:2005,Nagaosa:2025,Orignac:2003,Capelle:2007,Orgad:2001,Kuzmany:2008,Voit:1996,Dagotto:2024,Lebrat:2018}, a universal Luttinger liquid description of the attractive Fermi gas is rather rare \cite{Vincent:2008,Yang:2001,Shlyapnikov:2009}, and some fundamental questions still remain elusive.
	For instance, can we build a framework of TLL for the bound states?
	Can we address the backward scattering process in excitations of the spin mode with an energy gap?
	Consequently, developing a universal Luttinger liquid description of the attractive Fermi gas is highly desirable.
	
	The usual TLL theory for interacting electrons relies fundamentally on linearizing the spin and charge excitation spectra near the Fermi points, where spin-up and spin-down electrons share identical Fermi surfaces. In this framework, the magnetic field is treated as an effective chemical potential in the spin sector, while the Zeeman effect in the charge sector—also induced by the magnetic field—is typically neglected. Consequently, the standard TLL description is restricted to the unpolarized state or the weakly polarized regime, in which the density difference between the two spin species remains sufficiently small (see e.g. \cite{Giamarchi:2004} and references therein).
	Although Zhao and Liu \cite{Vincent:2008} explored a bosonization approach for 1D attractive Fermi gases with polarization, rigorous calculations of the effective interaction from higher-order operators and the sine-Gordon term remain missing.
For attractive Fermi gases, the polarization $p$ is critical: it not only characterizes spin imbalance but also drastically modifies the system’s low-energy excitations and correlation functions \cite{Shiba:1991}. 
In the spin-balanced case, spin-sector excitations are gapped, whereas charge excitations are gapless. Once polarization is present, both spin and charge excitations become gapless \cite{Xi-Wen Guan:2023,Bolech:2009}. Additionally, polarization drives notable many-body phenomena, such as quantum phase transitions \cite{Xi-Wen Guan:2007,Xi-Wen Guan:2011,Liao:Nature2010}, abnormal RG flows of the sine-Gordon term \cite{Matveev:2005}, and spatial oscillations of the pair correlation function \cite{Vincent:2008,Yang:2001}. These spatial modulations
are hallmarks of the Fulde-Ferrell-Larkin-Ovchinnikov(FFLO) pairing state. Backscattering between Fermi points of bound pairs and unpaired fermions produces a 1D analog of the FFLO state, revealing the microscopic origin of FFLO behavior \cite{Ferrell:1964,Ovchinnikov:1965}.
If the gap is small and for small polarization, the gap-gapless transition is described by a commensurate-incommensurate (C-IC) phase transition induced by the polarization \cite{Giamarchi:1988,Giamarchi:1996,Giamarchi:2004}. 
However, the usual TLL theory that neglects the effects of the magnetic field on the charge sector does not allow one to treat the scenario with large polarizations. 
The aim of this paper is to go beyond this description by incorporating magnetic field effects in both spin and charge sectors, and to extend the analysis to arbitrary polarization regimes.
	
	%
	In this paper, we formulate a Luttinger liquid theory for 1D attractive Fermi gases with finite polarization using the bosonization method.
	Via this effective Luttinger liquid theory, we analytically derive the low-energy effective Hamiltonian  in both the weak- and strong-binding regimes. We demonstrate the emergence of a two-component Luttinger liquid that depends strongly on the interaction strength and polarization in the weak-binding regime, and the paired and unpaired fermios exhibit charge-charge separation in the strong-binding regime.
	Using the effective Hamiltonian we constructed, we further compute the pair correlation function for the FFLO  pairing state.
	We also discuss experimental measurements of these dynamical correlation functions, based on the charge-charge separation mechanism for the FFLO pairing phase.
	Our results advance  experimental observations of the Luther-Emery liquid and charge-charge separation in 1D attractive Fermi gases \cite{Kafle:2025}.

	The paper outline is as follows: Sec. \ref{Section II} briefly introduces the theoretical framework of TLL theory, including the model Hamiltonian, exact solution, and notation for the bosonization formalism.
	Sec. \ref{Section.III} constructs the low-energy effective Hamiltonian of the polarized attractive Fermi gas for both weak and strong interaction regimes, and discusses the renormalization group (RG) flow of the sine-Gordon term.
	Using the effective Hamiltonian, we  calculate the pair correlation function of the FFLO-like state.
	Sec. \ref{Section IV} presents our conclusions and outlook.
	

	\section{ Yang-Gaudin Model And Bosonization}\label{Section II}
	
	We consider a 1D delta-function interacting ultracold Fermi gas prepared in two hyperfine sublevels denoted as $|\uparrow \rangle$ and $|\downarrow \rangle$.
	The Hamiltonian is given by
	\begin{eqnarray}
		H=-\frac{\hbar^2}{2m}\sum_{i=1}^{N}\frac{\partial^{2}}{\partial x_{i}^{2}}+2c\sum_{i,j=1}^{N}\delta(x_i-x_j)-hM-\mu N, \label{Hamiltonian}
	\end{eqnarray}
	with total particle number $N=N_\uparrow+N_\downarrow$ and magnetization $M=N_\uparrow-N_\downarrow$, where $N_\sigma$ ($\sigma = \uparrow,\,\downarrow$) is the number of particles in spin-$\sigma$ state.
Here $h$ and $\mu$ stand for the external magnetic field and chemical potential, respectively.
In our calculation, we adopt a unit system by setting $\hbar=2m=1$ for convenience, and we will focus on attractive interaction with $c<0$.
The interaction strength is  $c\equiv-2\hbar^2/ma_{1D}$, where  $a_{1D}$ is an  effective 1D scattering length \cite{Olshanii:1998}. 
It will be convenient to define the dimensionless interaction strength $\gamma \equiv c/n$, where $n=N/L$ is the total particle density. 
	In the absence of an external magnetic field,  the spin degrees of freedom have a finite gap $\Delta_\sigma\approx \frac{\hbar ^2 n^2 }{2m}\left( \frac{\gamma ^2}{2} -\frac{\pi^2 }{8} \right)$\cite{Xi-Wen Guan:2007}, which can be regarded as the energy scale for breaking the bound state. 
Hamiltonian Eq.~\eqref{Hamiltonian} is known as the  Yang-Gaudin model, first solved by Yang \cite{C.N.Yang:1967} and Gaudin  \cite{Gaudin:1967} using the Bethe Ansatz.
The universal  thermodynamic properties of the system can be derived from the thermodynamic Bethe Ansatz (TBA) equations \cite{Takahashi:1994,Lai:1973,Lai:1971,Xi-Wen Guan:2007,Xi-Wen Guan:2023}.
	%

	%
	%
	%
	%
	%

	
	Low-energy properties of one-dimensional (1D)  interacting Fermi gases are effectively described by Luttinger liquid theory \cite{Cazalilla:2011, Giamarchi:2004}.
	In this framework, low-energy excitations are described by bosonic collective modes.
	The first bosonization step is to linearize the dispersion relation near Fermi points, replacing the quadratic form $\epsilon(k)=\hbar^2k^2/2m$ by a linear approximation.
	This yields the Tomonaga-Luttinger model, in which fermionic excitations split into left- ($L$) and right($R$)-moving  modes with dispersions  $\epsilon_{L,R}=\hbar v_{F}(\mp k-k_{F})$, respectively.
	Here $v_F$ ($k_F$) is the Fermi velocity (wave vector).
	Such excitations of spin and charge sectors in the 1D polarized attractive Fermi gas have been  extensively studied in literature \cite{Fuchs:2004,Xi-Wen Guan:2023,Bolech:2009}.
 The fermionic field operators $\psi_{L,\sigma}$ and $\psi_{R,\sigma}$ can be represented in terms of the fields $\phi_{\sigma}$ and $\theta_{\sigma}$, which satisfy the bosonic commutation relations $\left[\phi_{\sigma}(x),\partial_x\theta_{\sigma\prime}(x^{\prime})\right]=i\pi\delta_{\sigma,\sigma^{\prime}}\delta_{x-x^{\prime}}$, using the Bose-Fermi identity\cite{Giamarchi:2004}
	\begin{eqnarray}\label{Fermi_Boson_identity}
		\psi_{r,\sigma}&=&\frac{U_{r,\sigma}}{\sqrt{2\pi\alpha}}e^{irk_{F,\sigma}x}e^{-i\left[r\phi_{\sigma}(x)-\theta_{\sigma}(x)\right]},
	\end{eqnarray}
	here $\sigma=\uparrow,\downarrow$ is the spin index, $\alpha$ is a short distance cutoff, $U_{\sigma,r}$ is the Klein factor, and $r=+1$ ($-1$) for right (left) moving fermions.
	
	In the absence of the magnetic field (i.e., $h=0$), the effective Bosonization Hamiltonian takes the celebrated spin-charge separated form, with the spin and charge excitations featuring distinct sound velocities.
In the present work, we focus on the effect of a finite polarization, which has received relatively less attention. A finite spin polarization drastically modifies the low-energy properties of the system. This is highly relevant to cold-atom experiments, where spin-$\uparrow$/$\downarrow$ populations can be independently controlled \cite{Gély:2025}. 
 For attractive interaction, we will discuss below the weak-binding ($\Delta_\sigma/(|c| n)  \ll1$) and strong-binding($\Delta_\sigma/(|c| n) \gg 1$) limits separately.
	
		%
		%
		%

		\section{Effective Hamiltonian of the 1D Spin-imbalanced attractive Fermi gas }\label{Section.III}

		\subsection{Weak-binding regime}

		\subsubsection{Effective Hamiltonian}
	 In this section, we derive an effective Hamiltonian for a 1D attractive Fermi gas subject to an external magnetic field. 
		This Hamiltonian yields subtly distinct pairing behaviors originating from spin imbalance within the system.
We first examine the regime of weak binding, where the binding energy of paired bound states is far lower than the relevant energy scale of interest, and fluctuations associated with BCS pairing prevail. 
The magnetic field induces a mismatch of the Fermi surfaces for the two spin components, hence $\delta k_F \equiv k_{F\uparrow}-k_{F\downarrow} \neq 0$ and $\delta v_F \equiv v_{F\uparrow}-v_{F\downarrow} \neq 0$ in general.
		After linearizing the spectrum near $\pm k_{F\uparrow(\downarrow)}$
		%
		%
		and rewriting fermionic field operators on the four branches $\psi_{r,\sigma}$ ($r=\pm1$; $\sigma=\uparrow,\downarrow$) in terms of bosonic ones using Eq.~\eqref{Fermi_Boson_identity}, the free Gaussian Hamiltonian is expressed as
		\begin{eqnarray}\label{Hamiltonian_free}
			H_{0}&=&\frac{1}{2\pi}\int dx\sum_{\sigma=\uparrow,\downarrow}v_{F,\sigma}\left[(\nabla\theta_{\sigma})^{2}+(\nabla\phi_{\sigma})^{2}\right].
		\end{eqnarray}
		

			%
			Using the language of the g-ology model \cite{Solyom:1979}, the interaction between particles on the same branch is described by the $g_4$ process;
			while the interaction between particles in the different branches are described by the $g_1$ (backward-scattering) and the $g_2$ (forward-scattering) processes.
			%
			%
			%
			The most general effective Hamiltonian, including both the free part Eq.\eqref{Hamiltonian_free} and all the interactions can be expressed as
			\begin{equation}\label{Hamiltonian_imbalance}
				\begin{split}
					H=\frac{1}{2\pi}\int &dx\Big[	\left(v_{F\uparrow}+\tilde{g}_{4\uparrow}+\tilde{g}_{2\uparrow}-\tilde{g}_{1\uparrow}\right)\left(\nabla\phi_\uparrow\right)^{2}\\
					&+\left(v_{F\uparrow}+\tilde{g}_{4\uparrow}-\tilde{g}_{2\uparrow}+\tilde{g}_{1\uparrow}\right)\left(\nabla\theta_{\uparrow}\right)^{2}\\
					&+\left(v_{F\downarrow}+\tilde{g}_{4\downarrow}+\tilde{g}_{2\downarrow}-\tilde{g}_{1\downarrow}\right)\left(\nabla\phi_{\downarrow}\right)^{2}\\
					&+\left(v_{F\downarrow}+\tilde{g}_{4\downarrow}-\tilde{g}_{2\downarrow}+\tilde{g}_{1\downarrow}\right)\left(\nabla\theta_{\downarrow}\right)^{2}\\
					&+2\left(\tilde{g}_{4\perp}+\tilde{g}_{2\perp}\right)\left(\nabla\phi_{\uparrow}\nabla\phi_{\downarrow}\right)\\
					&+2\left(\tilde{g}_{4\perp}-\tilde{g}_{2\perp}\right)\left(\nabla\theta_{\uparrow}\nabla\theta_{\downarrow}\right)\\
					&+\tilde{g}_{1\perp}\cos\left(2\phi_{\uparrow}-2\phi_{\downarrow}-2\delta k_{F}x\right)
					\Big],
				\end{split}
			\end{equation}
			where $\tilde{g}=g/2\pi$.
			More details about the derivation of the Hamiltonian Eq.~\eqref{Hamiltonian_imbalance} can be found in the appendix.
			Penc and S\'{o}lyom's work on the 1D repulsive Hubbard model \cite{Solyom:1993} and Zhao and Liu's work on 1D interacting Fermi gases \cite{Vincent:2008} show results similar to Eq.~\eqref{Hamiltonian_imbalance} except for the backward scattering (sine-Gordon) term.
			This term is irrelevant for low-energy repulsive interactions but relevant for attractive ones, as it may renormalize to infinity and cannot be neglected here\cite{Giamarchi:2004}.
			
			To express the Hamiltonian in terms of spin and charge degrees of freedom, we substitute the definitions of spin and charge bosonic operators $\phi_{c(s)} \equiv (\phi_\uparrow \pm \phi_\downarrow)/\sqrt{2}$ and $\theta_{c(s)} \equiv (\theta_\uparrow \pm \theta_\downarrow)/\sqrt{2}$ into Eq.~\eqref{Hamiltonian_imbalance}, and after a straightforward calculation, the Hamiltonian can be rewritten as
			\begin{eqnarray}\label{Hamiltonian_cs1}
				&\ &H=H_{c}+H_{s}+H_{cs}+H_{SG}
			\end{eqnarray}
			where
			\begin{eqnarray}\label{Hamiltonian_cs2}
				\begin{aligned}
					H_{cs}&=\frac{1}{2\pi}\int dx\ \left(\Delta_{cs1}\nabla\theta_{c}\nabla\theta_{s}+\Delta_{cs2}\nabla\phi_{c}\nabla\phi_{s}\right),\\
					H_{c,s}&=\frac{1}{2\pi}\int\left[u_{c,s}K_{c,s}(\nabla\theta_{c,s})^{2}+\frac{u_{c,s}}{K_{c,s}}(\nabla\phi_{c,s})^{2}\right]dx, \\
					H_{SG}&=\frac{g_{1\perp}}{2\pi^2\alpha^2}\int \cos\left(2\sqrt{2}\phi_{s}-2{\delta k_{F}}x\right)dx.
				\end{aligned}
			\end{eqnarray}
			Here, all the parameters $\Delta_{cs1(2)}$, $K_{c(s)}$, and $u_{c(s)}$ are related to the g-ology parameters (see appendix). Refs.~\cite{Vincent:2008, Solyom:1993} propose a way to extract these parameters by mapping the system to a repulsive Hubbard model in the low filling limit.
			
			The first line in Eq.~\eqref{Hamiltonian_cs2}, i.e., $H_{cs}$, describes the coupling between spin and charge degrees of freedom. It is characterized by two coefficients $\Delta_{cs1}, \Delta_{cs2} \propto \delta v_F$. For spin-balanced systems, we have $\delta v_F=0$, hence $H_{cs}$ vanishes, spin and charge degrees of freedom exhibit independent collective excitations and  propagate at distinct velocities -- the hallmark of spin-charge separation.

			The spin imbalance induced by the magnetic field fundamentally alters this paradigm.
			The emergent spin-charge coupling hybridizes the excitation spectra, which prevents the factorization into individual spin or charge quasiparticles. Note that this term is not taken into account in the usual commensurate-incommensurate description, which is thus restricted to the regime of zero or small polarization. 
			In order to describe this coupled excitation, we construct composite bosonic operators with the following canonical transformation:
			\begin{eqnarray}
				&&\left(\begin{matrix}
					\phi_{1}\\
					\phi_{2}\\
				\end{matrix}
				\right)=
				\left[\begin{matrix}
					\cos(\zeta) & \sin(\zeta)\\
					-\sin(\zeta) & \cos(\zeta)\\
				\end{matrix}\right]^{-1}
				\left(	\begin{matrix}
					\phi_{c}\\
					\phi_{s}\\
				\end{matrix}\right),\label{phi_12}\\
				&&\left(\begin{matrix}
					\theta_{1}\\
					\theta_{2}\\
				\end{matrix}\right)=	
				\left[\begin{matrix}
					\cos(\beta) & \sin(\beta)\\
					-\sin(\beta) & \cos(\beta)\\
				\end{matrix}\right]^{-1}
				\left(	\begin{matrix}
					\theta_{c}\\
					\theta_{s}\\
				\end{matrix}\right),\label{theta_12}
			\end{eqnarray}
			where $\zeta$ and $\beta$ are determined by:
			\[
			\tan(2\beta)=\frac{\Delta_{cs1}}{\left(u_sK_s-u_cK_c\right)}, \,\,\tan(2\zeta)=\frac{\Delta_{cs2}K_cK_s}{\left(u_s K_c-u_cK_s\right)}.\]
			%
			%
			In terms of these new bosonic field operators $\phi_{1(2)}$ and $\theta_{1(2)}$, the Hamiltonian can be rewritten as
			\begin{eqnarray}
				H&\!=\!&H_1+H_2+H_{SG},\label{Hamaltonian_perturbation}\\
				H_{n}&\!=\!&\frac{1}{2\pi}\int \!\left[u_{n}K_{n}(\nabla\theta_{n})^{2}+\frac{u_{n}}{K_{n}}(\nabla\phi_{n})^{2}\right]dx\,,\;n=1,2\nonumber
			\end{eqnarray}
			and $H_{SG}$ is same the sine-Gordon term as the last line of Eq.~\eqref{Hamiltonian_cs2} with $\phi_s =\cos\left(\zeta\right)\, \phi_{2}-\sin \left(\zeta\right) \,\phi_{1} $, which couples the two composite boson modes.
			In the above equations, the velocity $u_{1(2)}$ and the Luttinger parameters $K_{1(2)}$ are determined by
			\begin{eqnarray}\label{U_12_K_12}
				\frac{u_1}{K_1}&=&\frac{u_c}{K_c}\cos^{2}\left(\zeta\right)+\frac{u_s}{K_s}\sin^{2}\left(\zeta\right)\nonumber \\
				&&-\Delta_{cs2}\cos\left(\zeta\right)\sin\left(\zeta\right),\nonumber \\
				\frac{u_2}{K_2}&=&\frac{u_c}{K_c}\sin^{2}\left(\zeta\right)+\frac{u_s}{K_s}\cos^{2}\left(\zeta\right) \nonumber \\
				&&+\Delta_{cs2}\cos\left(\zeta\right)\sin\left(\zeta\right),   \\
				u_1K_1&=&u_cK_c\cos^{2}\left(\beta\right)+u_sK_s\sin^{2}\left(\beta\right)\nonumber \\
				&&-\Delta_{cs1}\cos\left(\beta\right)\sin\left(\beta\right), \nonumber \\
				u_2K_2&=&u_cK_c\sin^{2}\left(\beta\right)+u_sK_s\cos^{2}\left(\beta\right)\nonumber \\
				&&+\Delta_{cs1}\cos\left(\beta\right)\sin\left(\beta\right).\nonumber
			\end{eqnarray}

The effective Hamiltonian presented in Eq.~\eqref{Hamaltonian_perturbation} highlights a key property: the low-energy excitations of the weakly attractive spin-imbalanced Fermi gas take the form of collective density oscillations of the composite bosonic fields $\phi_1$ and $\phi_2$.
Crucially, these normal modes are defined via the  $SU(2)$ rotation specified in Eqs.~\eqref{phi_12} and \eqref{theta_12}, which encapsulate spin-charge coupling within the bosonic fields $\phi_1$ and $\phi_2$. 
This observation indicates that the low-energy spectrum comprises two normal modes that simultaneously carry charge and spin degrees of freedom, matching the excitations of the fractional state analyzed in Ref.~\cite{Lederer:2000}.
Accordingly, $\phi_1$ and $\phi_2$  may be interpreted as field operators for novel quasiparticles that encode low-energy excitations via spin-charge coupling.
Excitation spectra featuring this type of spin-charge mixing were reported alongside pair-breaking excitation spectra in Ref.~\cite{Xi-Wen Guan:2023}.					
			Moreover, the Fermi wave number mismatch $\delta k_F$, which is proportional to the polarization appears explicitly in the Sine-Gordon term $H_{SG}$. Similar to the usual commensurate-incommensurate phase transition 
			\cite{Giamarchi:2004,Giamarchi:1991,Giamarchi:1996}, $\delta k_F$ also induces a phase transition driven by the magnetic field, which we will discuss below.			
			
			\subsubsection{RG equation and relevant sine-Gordon term}

			For weak interaction ($|g_{1\perp}|\ll1$), we can treat $H_{SG}$ as a perturbation in Hamiltonian Eq.~\eqref{Hamaltonian_perturbation}.
			In order to  avoid presenting much  mathematical calculation,  here we just  summarize the key results of the RG analysis. More detailed calculation is  provided in the appendix.
			Consider a two-point correlation function  of the field operator $\phi_{1}(\bm{r})$
			\begin{eqnarray}
				R(\bm{r}_{1}-\bm{r}_{2})&=&\bigg\langle \hat{T}_{\tau} e^{i\sqrt{2}\phi_{1}(\bm{r}_{1})}e^{-i\sqrt{2}\phi_{1}(\bm{r}_{2})}\bigg\rangle_{H},
			\end{eqnarray}
			where $\bm{r}\equiv(x,u_0\tau)$ is the space-time coordinate in the imaginary time Heisenberg picture, $u_0$ is a velocity, and $\hat{T}_{\tau} $ is the time order operator.
			Expanding perturbatively  the correlation function up to the  second order,  we find that it exhibits a power-law form $R(\bm{r})\sim |\bm{r}| ^{-K_{eff}}$.
			Crucially, the exponent $K_{eff}$, which reflects the low-energy properties of the system, must remain independent of the cutoff parameter $\alpha$, which is an extra induced parameter in order to avoid a divergence.
			That is, if we scale the cutoff $\alpha=\alpha_{0}e^{l}$ to $\alpha_{0}e^{l+\Delta l}$, $K_{eff}$ should remain the same.
			This allows us to obtain the following RG equations:
			\begin{eqnarray}\label{renormalization_equation}
				\begin{aligned}
					\frac{dK_{1}(l)}{dl}&=-\frac{y_{1\perp}^{2}K^{2}_{1}\sin^{2}\left(\zeta\right)}{2}J_{0}(\bm{\epsilon}), \\
					\frac{dy_{1\perp}(l)}{dl}&=\left[2-2\left(\sin^{2}(\zeta)K_{1}+\cos^{2}(\zeta)K_{2}\right)\right]y_{1\perp},
				\end{aligned}
			\end{eqnarray}
			where $y_{1\perp}\equiv g_{1\perp}/(2\pi u_0)$, $\bm{\epsilon}\equiv2\delta k_{F}\alpha_{0}$, and $J_{0}(x)$ is the  zero-order Bessel function.
			The  oscillatory nature of this  function, induced by the real-space sharp cutoff \cite{Schulz:1983}, is unimportant here.
			Note that $J_{0}(x)\approx 1$ for $x<1$ and decays rapidly to zero for $x>1$, so $J_{0}(\bm{\epsilon})$  in Eq.\eqref{renormalization_equation} can be approximated by the unit step function
			$\theta(1-\bm{\epsilon})$. For $\bm{\epsilon}<1$ or $\delta k_F<1/(2\alpha_0)$, $ \theta(1-\bm{\epsilon})=1$,
			the renormalization equation resembles the spin-balanced case: $y_{1\perp}$ flows to 0 for repulsive interaction
			but to infinity for attractive ones \cite{Giamarchi:2004}.
			For $\bm{\epsilon}>1$ or $\delta k_F>1/(2\alpha_0)$, hence $ \theta(1-\bm{\epsilon})=0$, $y_{1\perp}$ always flows to 0 such that the sine-Gordon term is irrelevant
			\cite{Giamarchi:1988,Giamarchi:1991,Schulz:1983}, and the system is described by the two-component Luttinger liquid with $H=H_1+H_2$.
			This can be understood as follows: the original cosine term in $H_{SG}$ given by
			$ \cos \left( 2\sqrt{2}\phi_{s}-2\delta k_F x  \right)$
			is negligible for large $\delta k_F$ due to the rapid oscillation, while reducing to the spin-balanced case for small $\delta k_F$ \cite{Giamarchi:2004}.
			
The cutoff parameter $\alpha_0$ is typically taken as the mean distance between fermions, i.e., $\alpha_0=L/N = 1/n$. 
			Using $\delta k_F = \pi (n_\uparrow-n_\downarrow) = \pi p n$,  we define $p\equiv (n_\uparrow-n_\downarrow)/n$, which measures the spin imbalance. 
We can thus define a critical magnetic field $h_c = \Delta_\sigma$ below which the polarization is zero $p=0$. If $h$ is larger than $h_c$ but $p \ll \Delta_\sigma/(|c| n)$,  then the usual commensurate-incommensurate description applies 
\cite{Giamarchi:1988,Giamarchi:1991,Giamarchi:1996} with a relation $p \propto (h-h_c)^{1/2}$ and the spin part is described with an effective TLL parameter of $K_\sigma^{\rm eff} = 1/2$. 
Note that in this regime close to the critical point, the RG description cannot be used since it would be in the strong coupling regime \cite{Schulz:1980}.


Moving away from the critical point ($h \gg h_c$), we enter a regime for which the polarization is essentially proportional to the magnetic field. This is the regime that can be consistently described by the RG procedure of the present paper. Note that the presence of the finite polarization is consistent with the BCS-FFLO transition behavior predicted from Bethe Ansatz and Green's function methods \cite{Rainer:1994,Dupuis:1995,Xi-Wen Guan:2011,Xi-Wen Guan:2007}.

			\subsubsection{Pair Correlation Function}
			
			In the previous section, we proved that the 1D attractive Fermi gas can be described by a two-component Luttinger liquid in the FFLO-like (polarized) phase.
			In this section, we discuss the pair correlation function for the 1D  attractive Fermi gas, which has been studied by means of various methods, including the  Bethe Ansatz \cite{Orso:2007,Xi-Wen Guan:2007}, density matrix renormalization group (DMRG) \cite{Heidrich-Meisner:2007,Ueda:2008}, bosonization \cite{Vincent:2008,Yang:2001}, conformal field theory(CFT) \cite{X.W.Guan:2011}, and mean-field theory \cite{Huse:2007}.
			Those results clearly show  the hallmark of the FFLO state: the pair correlation function oscillates in coordinate space and exhibits a nonzero center-of-mass-momentum of the bound pair.
			In order to further verify the validity of the effective Hamiltonian we constructed in the previous section, we first calculate the pair correlation function in momentum space using the effective Hamiltonian Eq.~\eqref{Hamaltonian_perturbation}, but ignoring the sine-Gordon term $H_{SG}$ by assuming that $p \gg \Delta_\sigma/(|c| n)$, and then compare it with the results given by CFT.
				\begin{figure}[t]
				\centering
				\includegraphics[scale=0.6]{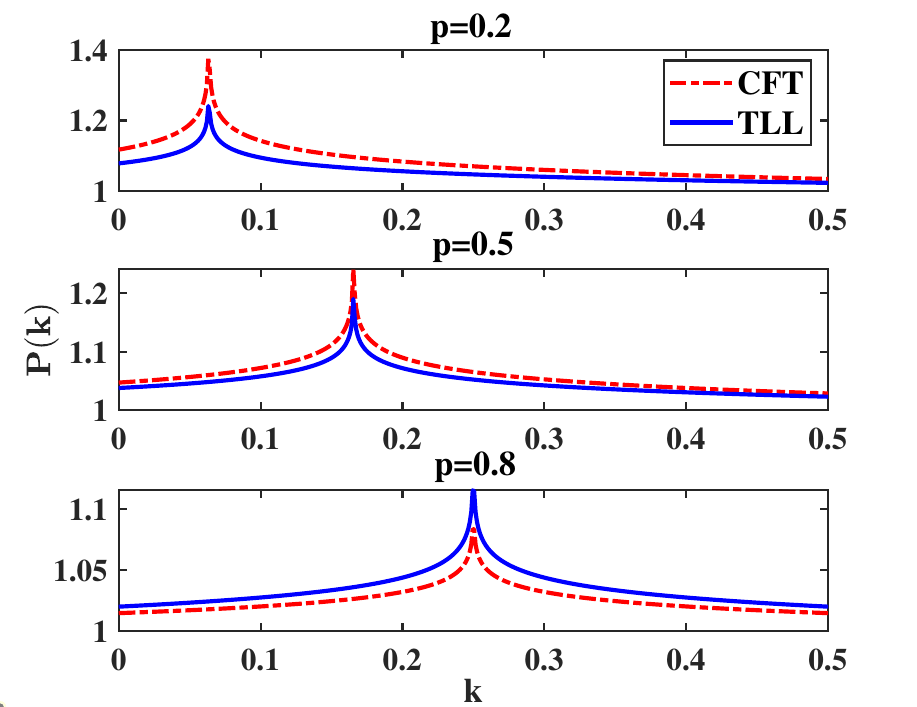}
				\caption{ Pair correlation function in momentum space. The result given by CFT(red danished line) predicted  in  \cite{X.W.Guan:2011} shares the common singular  feature   with  our  TLL results Eq.\eqref{pair_correlation_function_momentum_space}(blue solid line) for  different values of polarization $p=0.2,0.5,0.8$,  and the interaction strength $|\gamma|=10$. Here  the particle density  $n=0.1$.}
				\label{Fig3}
			\end{figure}
			The ground-state pair correlation in the coordinate space is defined by
			\begin{eqnarray}
				\mathcal{\bm{P}}(x,\tau)\equiv-\left\langle \hat{T}_{\tau} B^{\dagger}(x,\tau)B(0,0)\right\rangle_{GS},
			\end{eqnarray}
			where
			$
			B(x,\tau)=\psi_{L,\downarrow}(x,\tau)\psi_{R,\uparrow}(x,\tau)
			$
			is the annihilation operator of the bound pair.	
			Using the path integral method \cite{Giamarchi:2004}, we can analytically obtain
			\begin{eqnarray}\label{pair_correlation_function}
				\begin{split}
					\mathcal{\bm{P}}(x,\tau)&\sim\frac{e^{-i\delta k_Fx}}{\left(2\pi\alpha\right)^2}e^{-\eta_{2}F_{a}((x,\tau),u_2)}\\
					&\times e^{-2\cos(\zeta)\sin(\beta)F_{b}((x,\tau),u_2)}\\
					&\times e^{-\eta_{1}F_{a}((x,\tau),u_1)}e^{-2\cos(\beta)\sin(\zeta)F_{b}((x,\tau),u_1)}\\
					&\overset{\tau\rightarrow0}{\sim}\frac{e^{-i\delta k_{F}x}}{x^{\eta}},
				\end{split}
			\end{eqnarray}
			where $\eta=\eta_1+\eta_2$ with  $\eta_1=\sin^{2}(\zeta)K_1+\frac{\cos^{2}(\beta)}{K_1}, \eta_2=\cos^2(\zeta)K_2+\frac{\sin^{2}(\beta)}{K_2}$, and the functions $F_{a(b)}((x,\tau),u)$ are given in the appendix.
Eq.~\eqref{pair_correlation_function} clearly shows that the pair correlation function exhibits oscillation with power-law decay over distance,
			with the oscillation  wave vector $\delta k_F=k_{F \uparrow}-k_{F \downarrow}$.

The correlation functions in momentum space can be derived by taking the Fourier transform
			of $\mathcal{\bm{P}}(x,\tau)$. Using the method given in Ref.~\cite{Essler:2005}, the pair correlation function in momentum space at $k\simeq \delta k_F$ reads
			\begin{eqnarray}\label{pair_correlation_function_momentum_space}
				\begin{split}
					\mathcal{\bm{P}}(k)&\sim|k-\delta k_F|^{\nu}
				\end{split}
			\end{eqnarray}
			with  $\nu=1-\eta$.

On the other hand, the critical behavior of a system can be characterized via finite-size corrections to its low-lying excitations.
At zero temperature, the system’s correlation functions decay as power laws in spatial separation, whereas they exhibit exponential decay at finite temperatures.
The critical exponents governing universal two-point correlators of primary fields - such as the pair correlation function
$G_P(x,t)\sim \langle \psi ^\dagger _\uparrow (x,t)   \psi ^\dagger _\downarrow (x,t) \psi  _\uparrow (x,t)   \psi  _\downarrow (x,t) \rangle$, where   $\psi ^\dagger _\sigma$  and$\psi  _\sigma$
denote spin-$\sigma$ creation and annihilation operators, respectively - may be extracted from finite-size corrections of the underlying model.
Following standard CFT calculations, the asymptotic forms of two- or multi-point correlation functions can be evaluated directly from conformal dimensions. 
The latter quantities are solvable from the dressed energy equations, which serve as the Bethe ansatz counterparts in this framework.
The equal-time pair correlation functions near the singularity are given by \cite{X.W.Guan:2011}
$\tilde {G}_P (k)\sim [sign (k-\pi (n_\uparrow-n_\downarrow)] ^{2s_P }[k-\pi (n_\uparrow-n_\downarrow) ]^{\nu_P}$  with $s_P\approx 0$  and $\nu_P \approx -(1-P)/(2|\gamma|)$ for a strong attraction.

			Fig.~\ref{Fig3}  displays   comparison between the results given by TLL theory Eq.~\eqref{pair_correlation_function_momentum_space}  and the  CFT  result  \cite{X.W.Guan:2011}.

In our above  results,  the correlation exponent $\eta$  is determined by the parameters $K_{1(2)}, \zeta$ and $\beta$, which are defined in Eq.~\eqref{U_12_K_12}. 
The evaluation of $\eta$ thus requires the prior determination of $K_{c(s)},\, u_{c(s)}$, and $\Delta_{cs1(2)}$. 
For the attractive Fermi gas, however, these quantities are difficult to obtain directly. 
To circumvent this difficulty, Ref.~\cite{Vincent:2008, Solyom:1993} provides a method that maps the attractive Fermi gas into the repulsive Hubbard model in the low-filling limit; more details are provided in the appendix. 
Following this approach, we set the particle number density to $n=0.1$ and choose three different polarizations $p=0.2,0.5,0.8$, with interaction strength fixed at $c=-1$. 
It should be noted that although the corresponding coupling strength $|\gamma|=|c|/n=10$ falls within the strong-coupling regime, the binding energy of the bound pair $\epsilon=-\frac{1}{2}n_bc^2$ remains small. 
Although the singular behavior at the mismatch momenta $k= \pi (n_\uparrow-n_\downarrow)$ exhibits identical power-law decay with distance for both our bosonization approach and the CFT prediction \cite{X.W.Guan:2011}, their respective dynamical critical exponents differ slightly due to different methods.
In fact, the asymptotic correlation functions obtained via these distinct methods are only valid near the threshold.
Accordingly, the effective Hamiltonian given in Eq.~\eqref{Hamaltonian_perturbation} holds within this parameter regime.
We thus find a good consistency between the various approaches.
 In particular,  the method described in the present paper allows to reach regimes of sizable polarization usually out of reach for
the bosonization approach. 
On the other hand, when the polarization tends to zero, we do find that we go back smoothly to the exponents for the pair-pair correlation that would be predicted by the commensurate-incommensurate phase transition approach and the universal value of $K_\sigma^{\rm eff}=1/2$.

			\subsection{Strong binding regime}
			
\begin{figure}[t]
				\includegraphics[scale=0.55]{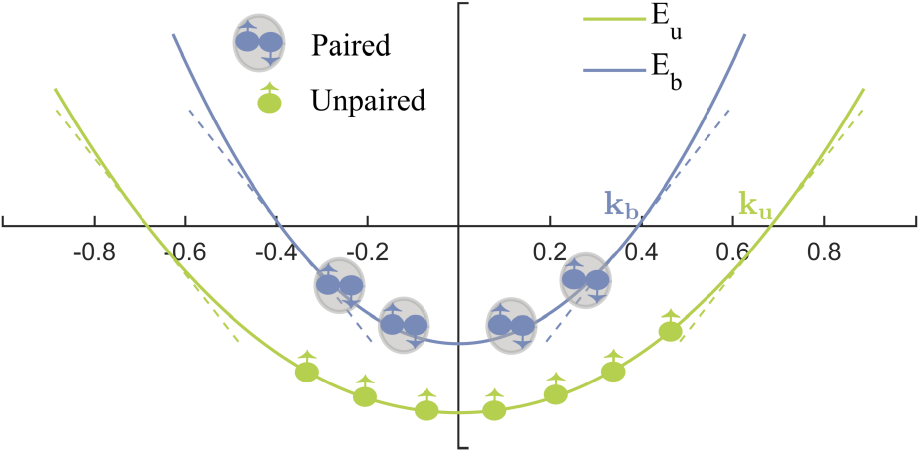}
				\caption{Energy spectra of the paired and unpaired fermions,  which are numerically calculated from the thermodynamic Bethe ansatz equations under the setting  $\mu=0,\, H=1,\,  |\gamma|=20$.
					The thermodynamic  Bethe ansatz equations are given in \cite{Xi-Wen Guan:2023}.
					In the unpaired fermions sector($E_u$), there only exist fermions with spin-up.
					In the the bound pair sector($E_b$), there exist  equal numbers of fermions with spin-up and spin-down.
					At the low-energy scale, we can linearize the spectra at the Fermi points $k_b$  and $k_u$ respectively.}
				\label{Fig9}
\end{figure}

We consider the strong coupling regime, i.e.,  $\Delta_\sigma \gg (|c| n)$.
Technically, the bosonization procedure outlined above is strictly valid only in the parameter regime where the binding energy of the bound state is small. 
Nevertheless, it can be extended to the opposite limit, where the paired fermions form deeply bound states with binding energy far exceeding the relevant low-energy scale, namely, the strong-coupling regime.
 In this scenario, the deeply bound pairs remain stable at low energies and effectively suppress pair-breaking excitations. 
 Consequently, the low-energy dynamics is governed primarily by particle-hole gapless excitations of both bound pairs and unpaired fermions. 
 The suppression of pair-breaking processes naturally gives rise to a decoupling of spin and charge degrees of freedom, allowing the effective Hamiltonian to be expressed in a spin-charge separated form in the strong-coupling limit.

Firstly, let us consider the ground state energy, from which we systematically construct the effective Hamiltonian using the Bethe ansatz equations of the model and renormalization group method.
In the FFLO-like phase, the ground-state energy of the strongly attractive Fermi gas is expressed as \cite{X.W.Guan:2012}
			\begin{eqnarray}
				E=-\frac{1}{2}n_bc^2+E_u+E_b+E_I,
			\end{eqnarray}	
			where $E_u=\pi^2n_u^3/3$ and $E_b=\pi^2n_b^3/3$ are kinetic energies of the unpaired and bound fermions, whose number density is given by $n_u$ and $n_b$, respectively.
			The first term $-\frac{1}{2}n_bc^2$ is the binding energy of bound pairs; and the last term
			\begin{equation}
				E_I=\frac{n_b^3\pi^2}{3} \frac{(2n_u+n_b)n}{|\gamma|} +\frac{n_u^3\pi^2}{3} \frac{8n_b n}{|\gamma |} +\mathcal{O}(\frac{1}{|\gamma|^2}),\label{Free-Gas}
			\end{equation}
			denotes the interacting energy between pairs and that between paired and unpaired fermions, and vanishes in the limit of $|\gamma| \rightarrow \infty$.
			This vanishing interaction implies the system's charge degrees of freedom decouple into two Fermi seas of paired and unpaired fermions \cite{Xi-Wen Guan:2007,Orso:2007,HuiHu:2007,Shlyapnikov:2009}, and consequently the Hamiltonian separates into two distinct components of Luttinger liquid:
			\begin{eqnarray}\label{Hamiltonian_strong_attractive}
				H=H_U+H_B,
			\end{eqnarray}
			where $H_U$ is the Hamiltonian of unpaired fermions, and $H_B$ that of the bound pairs.
			We can now bosonize $H_U$ and $H_B$ independently using the standard techniques. We linearize the spectra of paired and unpaired fermions at their respective Fermi points $k_{Fb}$ and  $k_{Fu}$ see Fig.(\ref{Fig9}).
			For unpaired fermions, low-energy ferromagnetic spin excitations are suppressed, so we only consider particle-hole excitations.
			$H_U$ thus takes the form of a standard Luttinger liquid.
			\begin{eqnarray}\label{Hamiltonian_unpaired}
				H_{U}&=&\frac{1}{2\pi}\int dx\left[u_{u}K_{u}\left(\nabla\theta_{u}(x)\right)^{2}+\frac{u_{u}}{K_{u}}\left(\nabla\phi_{u}(x)\right)^{2}\right]\n
				&+&\frac{h}{2\pi}\int dx\nabla\phi_{u}(x).
			\end{eqnarray}

			 On the other hand, for the bound pairs, the low energy effective Hamiltonian should contain a sine-Gordon term in the form $\sim g_{1\perp}\int \cos\left[2\sqrt{2}\phi^{(b)}_{s}(x)\right]dx$. For strong attractive interaction, however, $\phi^{(b)}_{s}(x)$ is pinned to zero in order to minimize the energy. 
Hence the sine-Gordon term can be neglected. Consequently,
			we have
			\begin{equation}
				H_B=\frac{1}{2 \pi} \int\left[u_b K_b\left(\nabla \theta_b(x)\right)^2+\frac{u_b}{K_b}\left(\nabla \phi_b(x)\right)^2\right] d x.\label{Hamiltonian_paired _fermions}
			\end{equation}

\begin{figure}[h]
				\centering
				\includegraphics[scale=0.6]{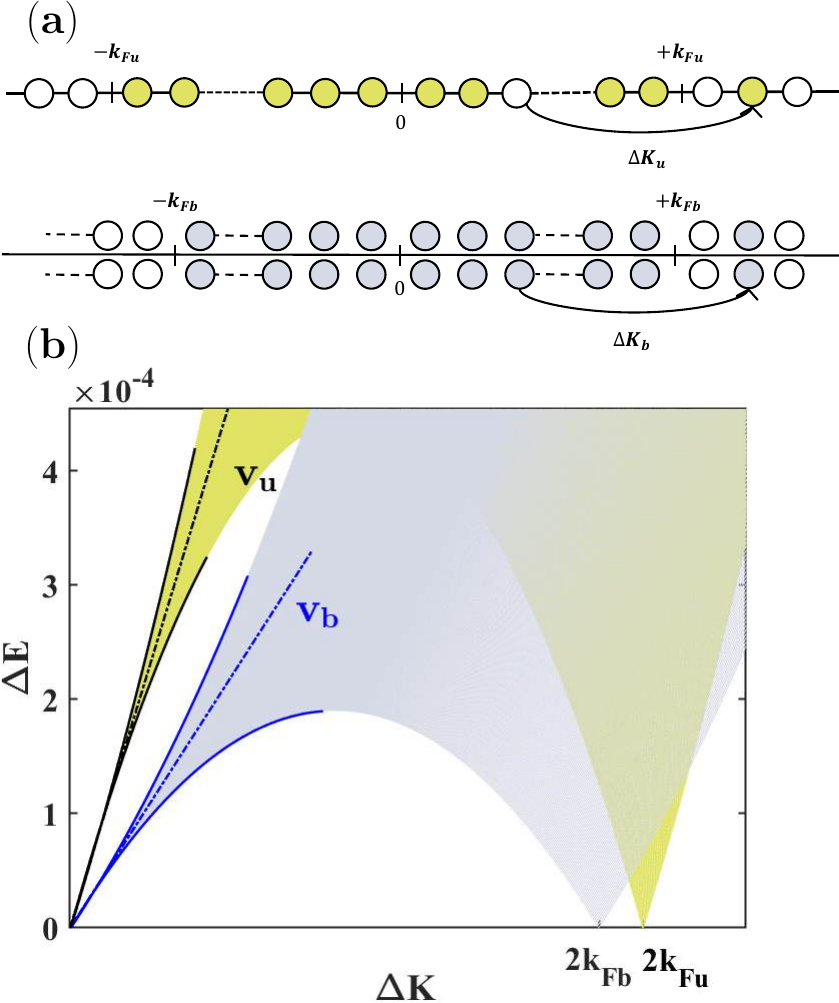}
				\caption{Exact low-energy excitations of the unpaired fermions and the bound pairs. (a) Schematic illustration of the particle-hole excitation of the unpaired fermions: moving one particle within the two Fermi points to outside the Fermi sea; the bound pairs: moving one pair of paired fermions within the two Fermi points to outside the Fermi sea without pair-breaking. (b) Particle-hole excitation spectra of the unpaired fermions(yellow green) and the bound pairs(blue) calculated from the Bethe ansatz equations, their spectra manifest a novel separation in their sound velocities, which is regarded as the charge-charge separation. This figure is calculated by solving the thermodynamic Bethe ansatz  equations numerically under the setting of the interaction strength $|\gamma |=52$ and polarization $p=0.35$. The thermodynamic  Bethe ansatz equations are given in \cite{Xi-Wen Guan:2023}. }
				\label{Fig4}
			\end{figure}

			%
			The sound velocity $u_{b(u)}$ and the Luttinger parameter $K_{b(u)}$ in $H_U$ and $H_B$ can be calculated analytically or numerically using the method of Bethe Ansatz \cite{X.W.Guan:2013}.
			\begin{eqnarray}
				\begin{aligned}
					u_{u}&\approx2\pi n_{u}\left[1+\frac{8n_{b}}{|c|}+\frac{48n_{b}^{2}}{c^{2}}\right], \\
					u_{b}&\approx\pi n_{b}\left[1+\frac{4n_{u}+2n_{b}}{|c|}+\frac{3(2n_{u}+n_{b})^{2}}{c^{2}}\right],\\
					K_{u}&\approx1+\frac{4n_{u}}{|c|}+\frac{4n_{u}(n_{u}+2n_{b})}{c^{2}}, \\
					K_{b}&\approx1+\frac{6n_{b}}{|c|}+\frac{3n_{b}(3n_{b}+4n)}{c^{2}}.
				\end{aligned}
			\end{eqnarray}
			In Fig.(\ref{Fig4}), we illustrate  the low-energy excitations of paired and unpaired fermions, from which sound velocities $u_b$ and $u_u$ are extracted in the long wave limit.
			Experimentally, this excitation spectrum and, hence the sound velocities, can be extracted using the technique of Bragg spectroscopy \cite{Hulet:2022}, from which the charge-charge separation phenomenon can be probed.

			\section{Conclusion and outlook}\label{Section IV}
			
			We have presented  an analytical  study of the Tomonaga-Luttinger liquid theory for the 1D attractive Fermi gas with both the weak and the strong binding regimes and arbitrary  polarizations.
			Based on bosonization, we have explicitly derived the low-energy effective Hamiltonian in the attractive Fermi gas across the weak and strong binding regimes, and further calculated the pair-correlation function for FFLO-like states. Our analysis, which takes into account a finite polarization, allows us to reach regimes of strong polarizion, something difficult to obtain in previous approaches, which were more concerned with the limit of either zero or small polarization. 
			We have found a description of the FFLO-like state in terms of  two-component TLL liquids,  exhibiting a spin-charge coupled regime in the weakly bound limit and charge-charge separation in the strongly bound limit.
		In particular, we have explicitly given  the renormalization group  equations for the sine-Gordon term in the spin sector and demonstrated that this term exhibits a behavior that is consistent  with the BCS-FFLO transition characteristics predicted by the Bethe Ansatz and Green’s function methods.

			Moreover, we further emphasize that the DSFs of the paired and unpaired fermions can be calculated analytically using the strong-binding effective Hamiltonian developed herein, with both quantities capturing characteristic signatures of the FFLO pairing state.

			 On the other hand,  those unpaired fermions with spin-up and bound pairs can be regarded as  free particles with different masses ($m$ and $2m$, respectively ), or in other words, without interaction between pairs and between a pair and a single atom  in the strong-coupling limit at low temperature, see Eq. (\ref{Free-Gas}).
			The   effective interactions of pairs and between pair and unpaired fermions can be taken into account in the effective sound velocities and effective masses of the paired and unpaired  atoms.
			Therefore, the paired and unpaired   DSFs can be approximated  by the expression \cite{Brand:2006,X.W.Guan:2020,Yang:PRL2018,Kafle:2025},
			namely,
			\begin{eqnarray}\label{DSFu_Free_fermion}
				S_{u,b} (q, \omega) =\frac{\operatorname{Im} \chi_{u,b}(q, \omega)}{\pi\left(1-e^{-\beta \hbar \omega}\right)},
			\end{eqnarray}
			Here $\chi_{u,b}(q,\omega)$ are the dynamical correlation functions of unpaired fermions and hard-core bound-pair particles, respectively; their imaginary parts with band curvature corrections are given explicitly in \cite{Kafle:2025}.
			Accordingly, paired and unpaired DSFs for attractively interacting  ${}^{6}$Li
			atoms at arbitrary polarizations can be measured via Feshbach resonance and Bragg spectroscopy, extending beyond the small-polarization regime of the recent experiment \cite{Kafle:2025}.

			

			\begin{acknowledgments}
				{\em  Acknowledgments--}
				X.W.G acknowledges support from the NSFC key grants No. 92365202, No. 12134015, U25D8013, and the Innovation Program for Quantum Science and Technology 2021ZD0302000.
				He is also partially supported by the Innovation Program for Quantum Science and Technology 2021ZD0302000.
				They acknowledge partial support from Quantum Science and Technology-National Science and Technology Major Project under Grant No. 2023ZD0300404 and the National Key R\&D Program of China under Grant No. 2022YFA1404104.
				HP is supported by the US NSF (Grant Nos. PHY-2513089 and the Welch
				Foundation (Grant No. C-1669). R.G.H. is supported by the US NSF (grant Nos. PHY-2309362 and PHY-2610079). This work is supported in part by the Swiss National Science Foundation under grant number 200020-219400.

			\end{acknowledgments}

			\appendix
			\setcounter{equation}{0}
			\renewcommand{\thefigure}{A\arabic{figure}}
			\renewcommand{\theequation}{A\arabic{equation}}
			
			\section{DATA AVAILABILITY}
			The numerical data presented in the figures are available at~\cite{Data}. Further raw data are also available from the author upon 
			request.
			
			\section{APPENDIX}
			In the appendix, we present additional details on analytical results from the main text.
			Sec.\ref{Section A} details the bosonization of the Hamiltonian in the weak coupling regime.
			Sec.\ref{Section B} gives the details of the determination of the parameters $K_{1(2)}, u_{1(2)}$ in the main text.
			Sec.\ref{Section C} derives the RG flow equations of the sine-Gordon term.

			\subsection{ The effective Hamiltonian from Bosonization }\label{Section A}
			In this part, we supplement in  details the bosonization  Hamiltonian.
			The free part of the Hamiltonian is expressed as a two-component Tomonaga-Luttinger model through  linearizing the spectrum at the Fermi points
			\begin{eqnarray}
				H_{0}&=&\frac{1}{2\pi}\int dx\sum_{\sigma=\uparrow,\downarrow}v_{F,\sigma}\left[(\nabla\theta_{\sigma})^{2}+(\nabla\phi_{\sigma})^{2}\right].
			\end{eqnarray}
			The interaction in  the effective Hamiltonian  is described by the $g_1,g_2, g_4$ process in  the language of the g-ology model.
			The $g_4$ process is expressed as
			\begin{eqnarray}\label{H_4}
				\begin{split}
					H_{4}&=H_{4||}+H_{4\perp}\\
					&=\int h_{4||}dx+\int h_{4\perp}dx\\
					&=\frac{g_{4\uparrow}}{(\pi)^{2}}\left[(\nabla\phi_{\uparrow})^{2}+(\nabla\theta_{\uparrow})^{2}\right]\\
					&\ \ +\frac{g_{4\downarrow}}{(2\pi)^{2}}\left[(\nabla\phi_{\downarrow})^{2}+(\nabla\theta_{\downarrow})^{2}\right]\\
					&\ \ +\frac{2g_{4\perp}}{(2\pi)^{2}}\left[\nabla\phi_{\uparrow}\nabla\phi_{\downarrow}+\nabla\theta_{\uparrow}\nabla\theta_{\downarrow}\right], \\[2mm]
				\end{split}
			\end{eqnarray}
			where
			\begin{eqnarray}   \label{h_4}
				\begin{split}
					h_{4||}(x)&=\sum_{r=R,L}\frac{g_{4\uparrow}}{2}\psi^{\uparrow\dagger}_{r,\uparrow}(x)\psi^{\uparrow\dagger}_{r,\uparrow}(x)\psi^{\uparrow}_{r,\uparrow}(x)\psi^{\uparrow}_{r,\uparrow}(x)\\
					&+\sum_{r=R,L}\frac{g_{4\downarrow}}{2}\psi^{\downarrow\dagger}_{r,\downarrow}(x)\psi^{\downarrow\dagger}_{r,\downarrow}(x)\psi^{\downarrow}_{r,\downarrow}(x)\psi^{\downarrow}_{r,\downarrow}(x)\label{6}, \\
					h_{4\perp}(x)&=\sum_{r=R,L}\frac{g_{4\perp}}{2}\psi^{\downarrow\dagger}_{r,\uparrow}(x)\psi^{\uparrow\dagger}_{r,\downarrow}(x)\psi^{\downarrow}_{r,\downarrow}(x)\psi^{\uparrow}_{r,\uparrow}(x)\\
					&+\sum_{r=R,L}\frac{g^{'}_{4\perp}}{2}\psi^{\downarrow\dagger}_{r,\uparrow}(x)\psi^{\uparrow\dagger}_{r,\downarrow}(x)\psi^{\downarrow}_{r,\downarrow}(x)\psi^{\uparrow}_{r,\uparrow}(x).
				\end{split}
			\end{eqnarray}
			Here the fermi filed $\psi_{r\sigma}^{\dagger\sigma^{'}}(x)\left(\psi_{r\sigma}^{\sigma^{'}}(x)\right)$
			means create (annihilate) a r(R or L) moving fermion with spin $\sigma(\uparrow \text{or} \downarrow)$, in the $\sigma^{'}(\uparrow \text{or} \downarrow)$ fermi surface.
			Using the Bose-Fermi identity Eq.~\eqref{Fermi_Boson_identity}, it can be expressed as the the form of bosonic operator.
			\begin{eqnarray}\label{Boson-Fermi identity}
				\psi_{r,\sigma}^{\sigma^{'}}&=&\frac{U_{r,\sigma}}{\sqrt{2\pi\alpha}}e^{irk_{F,\sigma}x}e^{-i\left[r\phi_{\sigma}(x)-\theta_{\sigma}(x)\right]}
			\end{eqnarray}
			for $\sigma^{'}=\uparrow$ or $\downarrow$.
			Taking  Eq.\eqref{Boson-Fermi identity} into Eq.\eqref{h_4}, and after a straightforward calculation we will get the Eq.\eqref{H_4}.
			The $g_2$ process is expressed by
			\begin{eqnarray}
				\begin{split}
					H_{2}&=H_{2||}+H_{2\perp}\\
					&=\int h_{2||}dx+\int h_{2\perp}dx\\
					&=\frac{g_{2\uparrow}}{(2\pi)^{2}}\left[(\nabla\phi_{\uparrow})^{2}-(\nabla\theta_{\uparrow})^{2}\right]\\
					&\ \ +\frac{g_{2\downarrow}}{(2\pi)^{2}}\left[(\nabla\phi_{\downarrow})^{2}-(\nabla\theta_{\downarrow})^{2}\right]\\
					&\ \  +\frac{2g_{2\perp}}{(2\pi)^{2}}\left[\nabla\phi_{\uparrow}\nabla\phi_{\downarrow}-\nabla\theta_{\uparrow}\nabla\theta_{\downarrow}\right].\\
				\end{split}
			\end{eqnarray}
			where
			\begin{eqnarray}
				\begin{split}
					h_{2||}(x)&=g_{2\uparrow}\psi^{\uparrow\dagger}_{L,\uparrow}(x)\psi^{\uparrow\dagger}_{R,\uparrow}(x)\psi^{\uparrow}_{R,\uparrow}(x)\psi^{\uparrow}_{L,\uparrow}(x)\\
					&+g_{2\downarrow}\psi^{\downarrow\dagger}_{L,\downarrow}(x)\psi^{\downarrow\dagger}_{R,\downarrow}(x)\psi^{\downarrow}_{R,\downarrow}(x)\psi^{\downarrow}_{L,\downarrow}(x), \label{13}\\
					h_{2\perp}(x)&=g_{2\perp}\psi^{\uparrow\dagger}_{L,\uparrow}(x)\psi^{\downarrow\dagger}_{R,\downarrow}(x)\psi^{\downarrow}_{R,\downarrow}(x)\psi^{\uparrow}_{L,\uparrow}(x)\\
					&+g_{2\perp}\psi^{\downarrow\dagger}_{L,\downarrow}(x)\psi^{\uparrow\dagger}_{R,\uparrow}(x)\psi^{\uparrow}_{R,\uparrow}(x)\psi^{\downarrow}_{L,\downarrow}(x).
				\end{split}
			\end{eqnarray}
			The $g_1$ process is given by
			\begin{eqnarray}
				\begin{split}
					H_{1}&=H_{1||}+H_{1\perp}=\int h_{1||}dx+\int h_{1\perp}dx,
				\end{split}
			\end{eqnarray}
			where
			\begin{eqnarray}
				\begin{split}
					h_{1||}(x)&=g_{1\uparrow}\psi^{\uparrow\dagger}_{L,\uparrow}(x)\psi^{\uparrow\dagger}_{R,\uparrow}(x)\psi^{\uparrow}_{L,\uparrow}(x)\psi^{\uparrow}_{R,\uparrow}(x)\\
					&+g_{1\downarrow}\psi^{\downarrow\dagger}_{L,\downarrow}(x)\psi^{\downarrow\dagger}_{R,\downarrow}(x)\psi^{\downarrow}_{L,\downarrow}(x)\psi^{\downarrow}_{R,\downarrow}(x), \label{16}\\
					h_{1\perp}(x)&=g_{1\perp}\psi^{\dagger\uparrow}_{L,\downarrow}(x)\psi^{\dagger\downarrow}_{R,\uparrow}(x)\psi^{\uparrow}_{L,\uparrow}(x)\psi^{\downarrow}_{R,\downarrow}(x)\\
					&+g_{1\perp}\psi^{\dagger\downarrow}_{L,\uparrow}(x)\psi^{\dagger\uparrow}_{R,\downarrow}(x)\psi^{\downarrow}_{L,\downarrow}(x)\psi^{\uparrow}_{R,\uparrow}(x).
				\end{split}
			\end{eqnarray}
			
			We find that $h_{1||}$ is identical to $h_{2||}$, but with a minus sign.
			Therefore we just need replace $g_{2\uparrow}$ and $g_{2\downarrow}$ by $(g_{2\uparrow}-g_{1\uparrow})$ and $(g_{2\downarrow}-g_{1\downarrow})$ when we calculate the total Hamiltonian.
			Furthermore  $h_{1\perp}$ can be expressed as
			\begin{eqnarray}
				\begin{split}
					h_{1\perp}&=\frac{g_{1\perp}}{(2\pi\alpha)^{2}}\left\{e^{2i[\phi_{\uparrow}-\phi_{\downarrow}-\delta k_{F}x]}+e^{-2i[\phi_{\uparrow}-\phi_{\downarrow}-\delta k_{F}x]}\right\}\\
					&=\frac{2g_{1\perp}}{(2\pi\alpha)^{2}}\cos\left(2\phi_{\uparrow}-2\phi_{\downarrow}-2\delta k_{F}x\right).
				\end{split}
			\end{eqnarray}
			The total  Hamiltonian Eq.\eqref{Hamiltonian_imbalance} includes  the free  Gaussian field  part and the interaction part.
			Summing up all of those, we obtain
			\begin{equation}
				\begin{split}
					H&=H_{0}+H_{\text{int}}\\
					&=H_{0}+H_{1}+H_{2}+H_{4}.
				\end{split}
			\end{equation}
			Further more,
			this effective Hamiltonian  can be given by
			\begin{eqnarray}
				\begin{split}
					H&=H_{c}+H_{s}\\
					&\ +\frac{1}{2\pi}\int dx\ \left(\Delta_{cs1}\nabla\theta_{c}\nabla\theta_{s}+\Delta_{cs2}\nabla\phi_{c}\nabla\phi_{s}\right)
				\end{split}
			\end{eqnarray}
			where
			\begin{eqnarray}
				\begin{split}
					H_{c}&=\frac{1}{2\pi}\int\left[u_{c}K_{c}(\nabla\theta_{c}(x))^{2}+\frac{u_{c}}{K_{c}}(\nabla\phi_{c}(x))^{2}\right]dx,\\
					H_{s}&=\frac{1}{2\pi}\int\left[u_{s}K_{s}(\nabla\theta_{s}(x))^{2}+\frac{u_{s}}{K_{s}}(\nabla\phi_{s}(x))^{2}\right]dx\\
					&\ 	+\frac{g_{1\perp}}{2\pi^2\alpha^2}\int cos\left[2\sqrt{2}(\phi_{s}-\frac{\delta k_{F}}{\sqrt{2}}x)\right]dx.
				\end{split}
			\end{eqnarray}
			
			The Luttinger parameters and the velocities are given by
			\begin{eqnarray}
				\begin{split}
					\Delta_{cs1}&=\delta v_{F}\left[1+\frac{\left(g_{4\uparrow}-g_{4\downarrow}\right)-\left(g_{2\uparrow}-g_{2\downarrow}\right)+(g_{1\uparrow}+g_{1\downarrow})}{2\pi\delta v_F}\right], \\
					\Delta_{cs2}&=\delta v_{F}\left[1+\frac{\left(g_{4\uparrow}-g_{4\downarrow}\right)+\left(g_{2\uparrow}-g_{2\downarrow}\right)-(g_{1\uparrow}-g_{1\downarrow})}{2\pi\delta v_F}\right], \\
					u_{c}K_{c}&=\bar{v}_{F}\Bigg[1+\frac{(g_{4\uparrow}+g_{4\downarrow})/2+(g_{4\perp}+g^{'}_{4\perp})/2}{2\pi \bar{v}_{F}}\\ &-\frac{(g_{2\uparrow}+g_{2\downarrow})/2+g_{2\perp}-(g_{1\uparrow}+g_{1\downarrow})/2}{2\pi \bar{v}_{F}}\Bigg], \\
					\frac{u_{c}}{K_{c}}&=\bar{v}_{F}\Bigg[1+\frac{(g_{4\uparrow}+g_{4\downarrow})/2+(g_{4\perp}+g^{'}_{4\perp})/2}{2\pi \bar{v}_{F}}\\
					&+\frac{(g_{2\uparrow}+g_{2\downarrow})/2+g_{2\perp}-(g_{1\uparrow}+g_{1\downarrow})/2}{2\pi \bar{v}_{F}}\Bigg], \\
					u_{s}K_{s}&=\bar{v}_{F}\Bigg[1+\frac{(g_{4\uparrow}+g_{4\downarrow})/2-(g_{4\perp}+g^{'}_{4\perp})/2}{2\pi \bar{v}_{F}}\\
					&-\frac{(g_{2\uparrow}+g_{2\downarrow})/2-g_{2\perp}-(g_{1\uparrow}+g_{1\downarrow})/2}{2\pi \bar{v}_{F}}\Bigg], \\
					\frac{u_{s}}{K_{s}}&=\bar{v}_{F}\Bigg[1+\frac{(g_{4\uparrow}+g_{4\downarrow})/2-(g_{4\perp}+g^{'}_{4\perp})/2}{2\pi \bar{v}_{F}}\\
					&+\frac{(g_{2\uparrow}+g_{2\downarrow})/2-g_{2\perp}-(g_{1\uparrow}+g_{1\downarrow})/2}{2\pi \bar{v}_{F}}\Bigg].
				\end{split}
			\end{eqnarray}

\subsection{ The determination of the parameters $K_{1(2)}, u_{1(2)}$}\label{Section B}
In this part, we calculate the parameters $K_{1(2)}$ and $u_{1(2)}$ for  the  effective Hamiltonian Eq.\eqref{Hamaltonian_perturbation} in the  weak binding regime. 
These parameters are determined  by the Eq.~\eqref{U_12_K_12} in the main text, and we need to compute the parameters $K_{c(s)}$ and $u_{c(s)}$ first.  
However, for the spin-charge coupling excitation, the linear spectra with sound velocity $u_c (u_s)$ in charge (spin) mode are covered  by the spin-charge coupled excitations, and the spectra are not linear at low-energy anymore.
Therefore,  we can not  obtain directly  the sound velocity of charge ($u_c$) and spin ($u_s$) mode.
 Meanwhile  the spin-charge coupling parameter $\Delta_{cs1(2)}$ is also difficult  to calculate.
  Ref.~\cite{Vincent:2008, Solyom:1993} provides a method to  calculate them,  by mapping the attractive Fermi gas to a corresponding repulsive Hubbard model in the limit of $n\ll1$. 
  Notably, the  Hamiltonian of the 1D Fermi gas Eq.~\eqref{Hamiltonian}  essentially relates to the Hamiltonian of 1D Hubbard model by the relations
\begin{eqnarray}
				m=\frac{\hbar^2}{2ta}\ \ \ c=-\frac{Ua}{2}\ \ \ \gamma=\frac{U}{2tn}, 
			\end{eqnarray}
where $t$  and $U$ denote the hopping parameter and the on-site interacting strength  in the Hamilton of the 1D Hubbard model
			\begin{eqnarray}
	H=-t\sum_{j,\sigma}\left(c^{\dagger}_{j,\sigma}c_{j+1,\sigma}+H.c\right)-U\sum_{j}n_{j\uparrow} n_{j\downarrow}.
\end{eqnarray}
Consequently, he Luttinger parameters $K_{c(s)}$ and the velocities $u_{c(s)}$  in Eq.\eqref{U_12_K_12} can be expressed as
			\begin{eqnarray}\label{U_cU_s_K_cK_s}
				\begin{split}
					u_cK_c&=\frac{u^{c}_{H}}{2}\left(\bar{Z}_{12}-\bar{Z}_{11}\right)^{2}+\frac{u^{s}_{H}}{2}\left(\bar{Z}_{22}-\bar{Z}_{21}\right)^{2}, \\
					u_sK_s&=\frac{u^{c}_{H}}{2}\left(\bar{Z}_{12}+\bar{Z}_{11}\right)^{2}+\frac{u^{s}_{H}}{2}\left(\bar{Z}_{22}+\bar{Z}_{21}\right)^{2}, \\
					\frac{u_c}{K_c}&=\frac{u^{c}_{H}}{2}\left(M_{12}-M_{11}\right)^{2}+\frac{u^{s}_{H}}{2}\left(M_{22}-M_{21}\right)^{2}, \\
					\frac{u_s}{K_s}&=\frac{u^{c}_{H}}{2}\left(M_{12}+M_{11}\right)^{2}+\frac{u^{s}_{H}}{2}\left(M_{22}+M_{21}\right)^{2}, \\
					\Delta_{cs1}&=u^{c}_{H}\left(\bar{Z}^{2}_{12}-\bar{Z}^{2}_{11}\right)+u^{s}_{H}\left(\bar{Z}_{22}^{2}-\bar{Z}^{2}_{21}\right), \\
					\Delta_{cs2}&=u^{c}_{H}\left(M^{2}_{12}-M^{2}_{11}\right)+u^{s}_{H}\left(M_{22}^{2}-M^{2}_{21}\right).
				\end{split}
			\end{eqnarray}
Here   $u^{c(s)}_{H}$ is the excitation sound velocity of the charge (spin)  mode of the corresponding repulsive Hubbard model \cite{Essler:2005}. 
$\bar{Z}$ and $M$ are defined as $M\equiv\left(\bar{Z}^\top\right)^{-1} $
			\begin{eqnarray}
				\bar{Z}\equiv
				\left[\begin{matrix}
					Z_{cc}-Z_{sc} & Z_{sc}\\
					Z_{ss}-Z_{cs} & -Z_{ss}\\
				\end{matrix}\right],
			\end{eqnarray}
			where $Z$ is the dress charge of the corresponding repulsive Hubbard model
			\begin{eqnarray}
				Z\equiv
				\left[\begin{matrix}
					Z_{cc} & Z_{cs}\\
					Z_{sc} & Z_{ss}\\
				\end{matrix}\right].
			\end{eqnarray}
			Once we get $u_{c(s)}$, $K_{c(s)}$ and $\Delta_{cs1(2)}$ according to Eq.~\eqref{U_cU_s_K_cK_s}, the parameters $K_{1(2)}$  and  $u_{1(2)}$ can be mathematically determined using Eq.\eqref{U_12_K_12}.

	\subsection{ RG equations for f the Sine-Gordon term}\label{Section C}
			In this part, we derive the RG equations  for the Sine-Gordon term, i.e., the Eq.\eqref{renormalization_equation}.
			Firstly, we introduce a formula, which is useful when we calculate the correlation function using the TLL theory.
			\begin{eqnarray}
				\begin{split}
					&\mathcal{I}=\left\langle\prod_{l}e^{i\left[A_{l}\phi(\bm{r}_{l})+B_{l}\theta(\bm{r}_{l})\right]}\right\rangle_{H_{TLL}}\n
					&\ =\exp\Biggl\{-\frac{1}{2}\sum_{m<l}\Big\{\left[-A_{m}A_{l}K-B_{m}B_{l}K^{-1}\right]F_{a}(\bm{r}_{m}-\bm{r}_{l},u)\\
					&\ +\left[A_{m}B_{l}+B_{m}A_{l}\right]F_{b}(\bm{r}_{m}-\bm{r}_{l},u)\Big\}\Biggr\}
				\end{split},
			\end{eqnarray}
			where $\mathcal{I}$ is the space-time correlation function of the low-energy excitation state of the Luttinger liquid $H_{TLL}$, where $\phi(\bf{r})$ is the bosonic field operator, $\mathbf{r}\equiv(x,y=u_0\tau)$ is the space-time coordinate of the 1D systems in the imaginary time Heisenberg picture and $u_0$ is a velocity.  The formula holds only if $\sum_{l}A_{l}=\sum_{l}B_{l}=0$, otherwise $\mathcal{I}=0$. The function $F_{a(b)}(\mathbf{r},u)$ is defined as
			\begin{eqnarray}
				F_{a}(\mathbf{r},u)&=&\frac{1}{2}\ln\left[\frac{x^{2}+(u|\tau|+\alpha)^{2}}{\alpha^{2}}\right], \\
				F_{b}(\mathbf{r},u)&=&-i\text{Sign}(\tau)\arctan\left[\frac{x}{u|\tau|+\alpha}\right].
			\end{eqnarray}
			We emphasize that the velocity $u$ in the function $F_{a(b)}$ is the sound velocity in the Hamiltonian of Luttinger liquid $H_{TLL}$ rather than the velocity $u_0$ in space-time coordinate.
			
			Now, we consider a two-point correlation function
			\begin{eqnarray}
				\mathcal{R}(\bm{r}_{1}-\bm{r}_{2})&=&\bigg\langle e^{i\sqrt{2}\phi_{1}(\bm{r}_{1})}e^{-i\sqrt{2}\phi_{1}(\bm{r}_{2})}\bigg\rangle_{H}.
			\end{eqnarray}
			Here, the correlation function $\mathcal{R}(\bm{r}_{1}-\bm{r}_{2})$ denotes the expectation value with respect to the Hamiltonian $H$ given in Eq. (9), which includes the two-component Luttinger liquid $H_0$ and the sine-Gordon term $H^{\prime}$
			\begin{eqnarray}
				\begin{aligned}
					H_{0}&=\frac{1}{2\pi}\int\left[u_{1}K_{1}(\nabla\theta_{1})^{2}+\frac{u_{1}}{K_{1}}(\nabla\phi_{1})^{2}\right]dx\\
					&+\frac{1}{2\pi}\int\left[u_{2}K_{2}(\nabla\theta_{b})^{2}+\frac{u_{2}}{K_{2}}(\nabla\phi_{2})^{2}\right]dx,
				\end{aligned}
			\end{eqnarray}
			and
			\begin{eqnarray}
				\begin{aligned}
					H^{'}=\frac{g_{1\perp}}{2\pi^2\alpha^2}\int &\cos\Big[2\sqrt{2}\Big(\cos(\zeta)\phi_{2}-\sin(\zeta)\phi_{1}\\
					&-\frac{\delta k_{F}}{\sqrt{2}}x\Big)\Big]dx,
				\end{aligned}
			\end{eqnarray}
			respectively.
			For weak interaction, i.e., $g_{1\perp}\ll1$, the sine-Gordon term can be regarded as a perturbation.
			Thus, we expand the correlation function perturbatively
			\begin{eqnarray}\label{correlation_perturbatively}
				\begin{split}
					R(\bm{r_{1}}-\bm{r_{2}})&=\bigg\langle e^{ia\sqrt{2}\phi_{a}(\bm{r_{1}})}e^{-ia\sqrt{2}\phi_{a}(\bm{r_{2}})}\bigg\rangle_{H}\\
					&=\frac{\bigg\langle e^{ia\sqrt{2}\phi_{a}(\bm{r_{1}})}e^{-ia\sqrt{2}\phi_{a}(\bm{r_{2}})}\hat{S}\bigg\rangle_{H_{0}}}{\langle\hat{S} \rangle},
				\end{split}
			\end{eqnarray}
			where
			\begin{eqnarray}
				\begin{aligned}
					\hat{S}&=\sum_{n}\frac{(-1)^{n}}{n!}\int_{0}^{\beta}d\tau_{1}\int_{0}^{\beta}d\tau_{2}\\
					&\cdot\cdot\cdot\int_{0}^{\beta}d\tau_{n}\hat{T}_{\tau}\left[H^{\prime}(\tau_{1})H^{\prime}(\tau_{2})\cdot\cdot\cdot H^{\prime}(\tau_{n})\right].
				\end{aligned}
			\end{eqnarray}
			We expand the correlation to second order, where
			\begin{eqnarray}\label{S_operator}
				\begin{split}
					\hat{S}_0&=1,\\
					\hat{S}_1&=-\int_{0}^{\beta}d\tau_{1}\ H^{\prime} \left(\tau_{1}\right),\\
					\hat{S}_2&=\frac{1}{2}\int_{0}^{\beta}d\tau_{1}\int_{0}^{\beta}d\tau_{2}\hat{T}_{\tau}\left[H^{\prime}  \left(\tau_{1}\right)H^{\prime} \left(\tau_{2}\right)\right].
				\end{split}
			\end{eqnarray}
			Substituting  Eq.\eqref{S_operator} into the Eq.\eqref{correlation_perturbatively}, we get
			\begin{equation}
				\begin{split}
					&\mathcal{R}(\bm{r}_{1}-\bm{r}_{2})=\bigg\langle e^{i\sqrt{2}\phi_{1}(\bm{r}_{1})}e^{-i\sqrt{2}\phi_{1}(\bm{r}_{2})}\bigg\rangle_{H_{_{0}}}\\
					&\ \ \ \ +\ \ \frac{1}{2}\frac{g_{1\perp}^{2}}{\left(2\pi\alpha\right)^{4}u_0^{2}}\int_{|\bm{r}|>\alpha}\ d^{2}\bm{r}^{'}\int d^{2}\bm{r}^{''}\\
					&\ \ \ \ \times \sum_{\epsilon_{1},\epsilon_{2}=\pm1}\Bigg\{\bigg\langle e^{i\sqrt{2}\phi_{1}(\bm{r}_{1})}e^{-i\sqrt{2}\phi_{1}(\bm{r}_{2})}\\
					&\ \ \ \ \times  e^{i\epsilon_{1}\sqrt{8}\left[\cos(\zeta)\phi_{2}(\bm{r}^{'})-\sin(\zeta)\phi_{1}(\bm{r}^{'})-\frac{\delta k_{F}}{\sqrt{2}}x^{'}\right]}\\
					&\ \ \ \ \times e^{-i\epsilon_{2}\sqrt{8}\left[\cos(\zeta)\phi_{2}(\bm{r}^{''})-\sin(\zeta)\phi_{1}(\bm{r}^{''})-\frac{\delta k_{F}}{\sqrt{2}}x^{''}\right]}\bigg\rangle_{H_{0}}\\
					&\ \ \ \ -\ \ \bigg\langle e^{ia\sqrt{2}\phi_{1}(\bm{r}_{1})}e^{-i\sqrt{2}\phi_{1}(\bm{r}_{2})}\bigg\rangle_{H_{0}}\\
					&\ \ \ \ \times \bigg\langle e^{i\epsilon_{1}\sqrt{8}\left[\cos(\zeta)\phi_{2}(\bm{r}^{'})-\sin(\zeta)\phi_{1}(\bm{r}^{'})-\frac{\delta k_{F}}{\sqrt{2}}x^{'}\right]}\\
					&\ \ \ \ e^{-i\epsilon_{2}\sqrt{8}\left[\cos(\zeta)\phi_{2}(\bm{r}^{''})-\sin(\zeta)\phi_{1}(\bm{r}^{''})-\frac{\delta k_{F}}{\sqrt{2}}x^{''}\right]}\bigg\rangle_{H_{0}}\Bigg\},
				\end{split}
			\end{equation}
			where we denoted $d^{2}\bm{r}=dxdy$.
			We  now introduce the center of mass and relative coordinates
			\begin{eqnarray}
				x=x'-x''\ \ \bm{R}=\frac{\bm{r}^{'}+\bm{r}^{''}}{2}\ \ \bm{r}=\bm{r}^{'}-\bm{r}^{''},
			\end{eqnarray}
			then the  correlation function can be rewritten as
			\begin{equation}
				\begin{split}
					&\mathcal{R}(\bm{r}_{1}-\bm{r}_{2})=e^{-K_{1}F_{a}(\bm{r}_{1}-\bm{r}_{2},u_1)}\\
					&\ \ \times \Bigg\{1+\frac{1}{2}\frac{g_{1\perp}^{2}}{\left(2\pi\alpha\right)^{4}u_1^{2}}\sum_{\epsilon_{1}=\pm1}\int_{|\bm{r}|>\alpha} d^{2}\bm{r}\int d^{2}\bm{R} \\
					&\ \  \times   e^{-4\sin^{2}(\zeta)K_{1}F_{a}(\bm{r},u_1)-4\cos^{2}(\zeta)K_{2}F_{a}(\bm{r},u_2)}\\
					&\ \  \times \left[e^{2\epsilon_{1}\sin(\zeta)K_{1}\big(\bm{r}\bm{\cdot\nabla_{R}}\left[F_{a}(\bm{r}_{1}-\bm{R},u_1)-F_{a}(\bm{r}_{2}-\bm{R},u_1)\right]\big)}-1\right]\\
					&\ \  \times e^{-i\epsilon_{1}2\delta k_{F}x}\Bigg\}.
				\end{split}
			\end{equation}
			We notice that the factor $$e^{-4\sin^{2}(\zeta)K_{1}F_{a}(\bm{r},u_1)-4\cos^{2}(\zeta)K_{2}F_{a}(\bm{r},u_2)}$$ is proportion to $\frac{1}{|\bm{r}|^2}$, and the integral over $\bm{r}$ will be dominant by the small part of $|\bm{r}|$. Thus, we can take $u_2$ approximately as $u_1$.
			Then, this factor can be written as $e^{-4\left[\sin^{2}(\zeta)K_{1}+\cos^{2}(\zeta)K_{2}\right]F_{a}(\bm{r},u_1)}$.
			For a small $|\bm{r}|$ in the integral, we can expand the exponential term in powers of $\bm{r}$. Then, we get
			
			\begin{equation}\label{A25}
				\begin{split}
					&\mathcal{R}(\bm{r}_{1}-\bm{r}_{2})=e^{-K_{1}F_{a}(\bm{r}_{1}-\bm{r}_{2},u_1)}\\
					&\ \ \ \ \times \Bigg\{1+\frac{g_{1\perp}^{2}\sin^{2}(\zeta)K_{1}^{2}}{8\pi^{4}\alpha^{4}u_0^{2}}\int_{|\bm{r}|>\alpha} d^{2}\bm{r}\int d^{2}\bm{R}\\
					&\ \ \ \ \times  e^{-4\left[\sin^{2}(\zeta)K_{1}+\cos^{2}(\zeta)K_{2}\right]F_{a}(\bm{r},u_1)}\\
					&\ \ \ \ \times\cos\left(2\delta k_{F}x\right)\big(\bm{r}\cdot\bm{\nabla_{R}}\big[F_{a}(\bm{r}_{1}-\bm{R},u_1)\\
					&\ \ \ \ - F_{a}(\bm{r}_{2}-\bm{R},u_1)\big]\big)^{2}\Bigg\}.\\
				\end{split}
			\end{equation}
			After a straightforward calculation with  the integral over $\bm{R}$, we may obtain
			\begin{equation}
				\begin{split}
					&\mathcal{R}(\bm{r}_{1}-\bm{r}_{2})=e^{-K_{1}F_{a}(\bm{r}_{1}-\bm{r}_{2},u_1)}\\
					&\ \ \ \ \times \bigg\{1+\frac{g_{1\perp}^{2}\sin^{2}(\zeta)K_{1}^{2}}{8\pi^{4}\alpha^{4}u_0^{2}}\\
					&\ \ \ \ \times \left[J_{+}I_{+}(r_{1}-r_{2})+J_{-}I_{-}(r_{1}-r_{2})\right]\bigg\},
				\end{split}
			\end{equation}
			where
			\begin{eqnarray}\label{A27}
				\begin{aligned}
					J_{\pm}&=\int_{r>\alpha} \frac{d^{2}\bm{r}}{\alpha^{4}}\left(x^{2}\pm y^{2}\right)\cos\left(2\delta k_{F}x\right)\\
					&\ \ \times e^{-4\left[\sin^{2}(\zeta)K_{1}+\cos^{2}(\zeta)K_{2}\right]F_{a}(\bm{r},u_1)}, \\
					I_{\pm}&=\int d^{2}\bm{R}F_{a}(\bm{r}_{1}-\bm{R},u_1)\left(\bm{\nabla}_{X}^{2}\pm\bm{\nabla}_{Y}^{2}\right)\\ &\ \ \times F_{a}(\bm{r}_{2}-\bm{R},u_1).
				\end{aligned}
			\end{eqnarray}
			Now, we should replace $F_{a}(\bm{r},u_1)$ by $F(\bm{r},u_1)$:
			\begin{eqnarray}\label{A28}
				F(\bm{r},u_1)=\frac{1}{2}\ln\left[\frac{x^{2}+(u_1|\tau|+\alpha)^{2}}{\alpha^{2}}\right]+\frac{t^{\prime}}{K_{a}}\cos\left(2\theta_{r}\right). \n
			\end{eqnarray}
			As we can see in Eq.\eqref{A25}, the symmetry between space and time is destroyed by the polarization $\delta{k_{F}}$.
			We introduce the parameter $t^{\prime} $  to  parameterize the anisotropy between space and time directions.
			The $t^{\prime}$  is zero in original free effective  Hamiltonian and nonzero  in term of the renormalization  process due to mismatch between the spin-up and -down Fermi surfaces.
			$\theta_{r}$  is the angle between vector $(x,y)$ and $x$-axis.
			Using eq.\eqref{A28}, we have
			\begin{eqnarray}
				\begin{aligned}\label{A29}
					J_{+}&=2\pi\int^{+\infty}_{\alpha}\frac{dr}{\alpha}\left(\frac{r}{\alpha}\right)^{3-4\left[\sin^{2}(\zeta)K_{a}+\cos^{2}(\zeta)K_{b}\right]}\\
					&\ \times J_{0}\left(2\delta k_{F}r\right), \\
					J_{-}&=2\pi\int^{+\infty}_{\alpha}\frac{dr}{\alpha}\left(\frac{r}{\alpha}\right)^{3-4\left[\sin^{2}(\zeta)K_{a}+\cos^{2}(\zeta)K_{b}\right]}\\
					&\ \times J_{2}\left(2\delta k_{F}r\right), \\
					I_{+}&=\pi \ln\left(\frac{r_{1}-r_{2}}{\alpha}\right), \\
					I_{-}&=\pi\cos\left(2\theta_{r_{1}-r_{2}}\right),
				\end{aligned}
			\end{eqnarray}
			where $J_{0}$ and $J_{2}$ is the zero and second order Bessel function.
			Here, we use the relation $d^{2}\bm{r}=2\pi |r|d|r|$ in the first two equation in Eq.\eqref{A29},  and $r\equiv|\bf{r}|$ is a scalar and no longer a 2D  space-time vector.
			In the last two equation in Eq.\eqref{A29}, we used the formula $\left(\bm{\nabla}_{x}^{2}\pm\bm{\nabla}_{y}^{2}\right)\ln(|\bm{r}|)=2\pi\delta(|\bm{r}|)$.
			Now, we introduce $K_{eff}$ and $t_{eff}$
			\begin{eqnarray}
				\begin{aligned}
					K_{eff}&=K_{1}-\frac{y_{\perp}^{2}\sin^{2}\left(\zeta\right)K^{2}_{1}}{2}\int^{+\infty}_{\alpha}\frac{dr}{\alpha}\\
					&\ \ \times \left(\frac{r}{\alpha}\right)^{3-4\left[\sin^{2}(\zeta)K_{1}+\cos^{2}(\zeta)K_{2}\right]}J_{0}\left(2\delta k_{F}r\right), \\
					t_{eff}&=t^{\prime} +\frac{y_{\perp}^{2}\sin^{2}\left(\zeta\right)K^{2}_{1}}{4}\int^{+\infty}_{\alpha}\frac{dr}{\alpha}\\
					&\ \ \times \left(\frac{r}{\alpha}\right)^{3-4\left[\sin^{2}(\zeta)K_{1}+\cos^{2}(\zeta)K_{2}\right]}J_{2}\left(2\delta k_{F}r\right)
				\end{aligned}
			\end{eqnarray}
			with  $y_{1\perp}=\frac{g_{1\perp}}{2\pi u_0}$,  the correlation function $\mathcal{R}(\bm{r}_1-\bm{r}_2)$ can be expressed as
			\begin{eqnarray}
				\mathcal{R}(\bm{r}_{1}-\bm{r}_{2})=e^{-\ln\left(\frac{r_{1}-r_{2}}{\alpha}\right)K_{eff}-\cos\left(2\theta_{\bm{r}_{1}-\bm{r}_{2}}\right)t_{eff}}. \n
			\end{eqnarray}
			
			It is well understood that  the exponents of the above correlation function  are determined by the interaction and the low energy excitations in the system.
			Therefore it  is  not relevant with the cut off parameter $\alpha$.
			If we rescale $\alpha=\alpha_{0}e^{l}$ to $\alpha^{'}=\alpha_{0}e^{l+dl}$, the effective exponent $K_{eff}$ should remain invariant.
			According to  this feature, we have  the following  RG equation
			\begin{eqnarray}
				\begin{aligned}
					\frac{dK_{1}(l)}{dl}&=-\frac{y_{1\perp}^{2}K^{2}_{1}\sin^{2}\left(\zeta\right)}{2}J_{0}\left(2\delta k_{F}\alpha_{0}\right), \\
					\frac{dy_{1\perp}}{dl}&=\left[2-2\left(\sin^{2}(\zeta)K_{1}+\cos^{2}(\zeta)K_{2}\right)\right]y_{1\perp}(l).
				\end{aligned}
			\end{eqnarray}

	\bibliographystyle{apsrev4-2}
			
		\end{document}